# Self-dispersion of Two Natural Polysaccharides for Granular Composites

Herbert Wang[1,2,*], Yin Fang[1], Yiliang Lin[1], Jiuyun Shi[1]

[1] The James Franck Institute, University of Chicago, Chicago, IL 60637.

[2] Hinsdale Central High School, Hinsdale, IL 60521.

*Correspondence to: herbertwang16@gmail.com

## Abstract

We envision that dispersion between two polymeric materials on mesoscales would create new composites with properties that are much more superior to the components alone. Here we elucidate the dispersion between two of most abundant natural polysaccharides, starch and chitosan, which form mesoscale composites that may promise many applications. By using X-ray microscopic imaging, small-angle X-ray scattering, and differential scanning calorimetry, we were able to characterize the interactions of chitosan and starch in the mesoscale composites. The morphology of the composite is far more complex from the simple mixture of starch granules with a nominal size of a few micrometers and chitosan microbundles of tens and hundreds of micrometers. This unique morphology can only be explained by the enhanced miscibility of chitosan in a starch granular matrix. It is evidenced that there is a possible ionic interaction between the amino group in chitosan and the hydroxyl groups in starch granules. Despite the limited solubility of chitosan in water, this ionic interaction allows for the disassembly of chitosan microbundles within the starch suspension. The result is a chemically stronger and more stable granular composite formed by two biocompatible and biodegradable polysaccharide polymers. The mechanism of chitosan to disperse throughout starch granules has implications for the application of chitosan in water and other solvents.

Polymer composites have been of intense industrial, technological, and scientific interest for decades. These composites are unique in their mesoscale interactions among their components [1]. For example, the mesoscale interactions between granules, such as starch particles, and other micro- and nano-sized particles allow for composite properties distinctive from those of each component itself [2]. The composites can exhibit improved tensile strength and material stability [3]. Due to the environmental problems on a global scale, the materials used to synthesize composites are sought to be biocompatible and biodegradable [4], while possessing unique biological, chemical, and physical properties.

Chitosan, α (1→4)-linked 2-amino-2-deoxyβ-D-glucopyranose (shown in **Figure 1a**), is the deacetylated form of chitin, which is an abundant polysaccharide and biopolymer found in nature. Chitin is found in the exoskeletons of arthropods and is most commonly acquired from the shells of crustaceans. Chitosan has been commonly utilized for its properties like biocompatibility, non-toxicity, antimicrobial and fungicidal abilities, and antioxidativity [5, 6]. Chitosan is the most common *N*-deacetylation derivative of chitin, formed at a degree of acetylation of less than 35% [7]. *N*-deacetylation removes the acetyl groups off of the molecular backbone, and the degree of deacetylation affects chitosan's molecular weight, crystallinity, and hydrophilicity, all factors that have an impact on chitosan's solubility [8, 9]. The cohesive energy density of chitosan is relatively high, due to its semi-crystalline structure and large amounts of hydrogen bonds, which makes it insoluble in water [10]. Chitosan has poor solubility in other agents too, as it is only slightly soluble (< 2%) in weak acidic solutions with a pH < 6, facilitated by the interactions with the charged amino group of chitosan; however, when the pH changes back to 6, the chitosan starts to precipitate [11]. This poor solubility makes it difficult to be effectively utilized as a functional biomaterial when it is mixed in many inorganic and organic materials. Thus, methods have been developed to increase chitosan's solubility in water, such as decreasing chitosan's molecular weight in order to reduce the crystallinity [12], chemically modifying the amino group using hydrophilic moieties [13], or adding charge groups on the chitosan backbone [14]. These processes are complex and can adversely impact chitosan's intrinsic properties. Consequently, a simpler and more compatible solution for improving chitosan solvation needs to be discovered and developed.

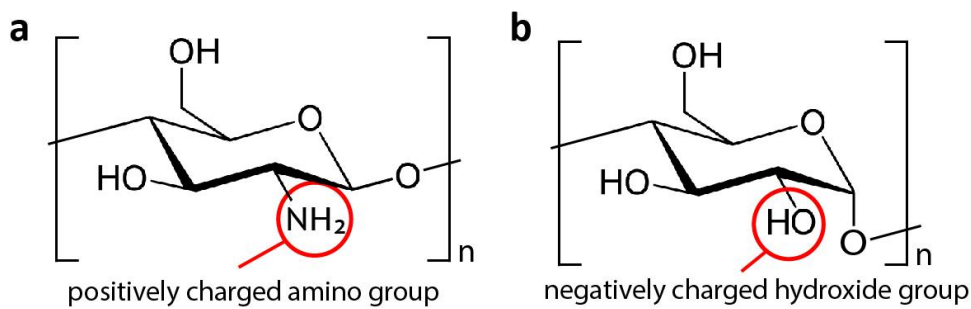

**Figure 1. Chitosan and starch have similar monomer structure but carry opposite charges. a.** Chitosan **b.** Starch (amylose)

On the other hand, starch is also one of the most abundant polysaccharides found in nature as a source of energy for plants, thus making it extremely low-cost to produce and use. Starch, like chitosan, is biocompatible and has excellent processability and renewability. In addition, it is biodegradable and has tough mechanical strength [15]. Starch structure varies from one kind to another as each has a varying size from 1 µm to more than 100 µm, a varying shape from ovals to half-spheres, and varying compositions of amylose and amylopectin [16]. Despite intensive research, the exact hierarchical structure of the granular particles is still elusive. It is known that starch's granular structure is dependent on the composition of amylose and amylopectin, α-1-4 linked glucopyranose with α-1-6 branches. Cereal grain starches like wheat and rice starches display an A-pattern diffraction pattern revealing the semi-crystalline nature of starch [17]. Amylopectin forms a branched structure, with a common crystalline model being a spherocrystalline assembly of disk-shaped amylopectin molecules radially facing outwards [18]. This standard model characterizes amylose as usually in a single and double helix structure and being stacked onto the amylopectin branches, which forms a network of crystalline and amorphous structure [19, 20]. Because of the dense arrangement of compact molecules and semi-crystalline structure, starch is water-insoluble. While insoluble, the granules of starch, of micrometer size, can be dispersed in water in a suspension.

One of starch's components in the polymer structure, amylose, has a similar structure to chitosan, except each monomer contains negatively charged hydroxyl groups (**Figure 1b**) in comparison to chitosan's positively charged amino groups (**Figure 1a**). These charged groups could result in polyelectrolyte bonding between the two saccharides, possibly forming ionic interactions. Herein, we envision to create a suspension of chitosan and starch in order to enable the dispersion of chitosan among starch granules. We utilized rice starch due to its small nominal size of around 2 to 7 µm and its superior monodispersity properties compared to other starches. Not only will the proposed method of dispersing chitosan be low-cost and simple, but it also preserves chitosan's intrinsic functionality as well as providing structural benefits. Using a host of structural and functional characterizing methods on many length scales such as X-ray microscopic imaging, small-angle X-ray scattering (SAXS), and differential scanning calorimetry (DSC), we elucidated the dissembling of chitosan microbundles possibly into nanofibers, which facilitates the dispersion of the two materials in a water suspension and the creation of a biocomposite.

In the experiments, we created suspensions of pure chitosan, pure starch (all purchased from Sigma Aldrich), and starch-chitosan with varying stoichiometries. We focused on the samples with the compositions 12/88 (wt%, chitosan/water) for the pure chitosan sample, 31/69 (wt%, starch/water) for the pure starch sample, and 29/9/62 (wt%, starch/chitosan/water). The chitosan and starch were stirred in deionized water for 48 hours using a magnetic stir bar with a stabilized temperature at 25 °C.

X-ray microscopic imaging was performed at beamlines 7-ID-B and 8-ID-E beamlines using either polychromatic (7-ID-B) and monochromatic (8-ID-E) of the Advanced Photon Source (APS), Argonne National Laboratory. Unlike conventional radiography that records only the absorption of the samples, the phase-contrast images also record phase information due to weak scattering, which enhances the contrast of weakly absorbing materials [21] like polysaccharide-based carbohydrates. The phase-contrast provided clarity to the images of starch granules and chitosan microbundles. Due to the short wavelength of around 0.1 nm, X-ray scattering can be used to understand the structure of materials on nanoscales. SAXS patterns, recorded as a function of the scattering angle in the forward direction (typically in the range of 0.1 to several degrees), was used to show the structural evolution of the starch-chitosan suspension on nanometer length scale. The measurement was performed at beamline 12-ID-B pf the APS. Gel samples were prepared and trimmed into long strips with a 2 × 2 mm$^2$ cross-section and mounted to a home-made sample holder so that the incident X-rays were perpendicular to the long axis of the strips. The wavelength (λ), size, and flux of the incident X-ray beam were 0.093 nm, 0.5 × 0.5 mm$^2$, and 1 × 10$^{12}$ photons/s. A Pilatus 2M detector, placed at 2 m from the samples, was used to record the 2D scattered X-ray patterns. The 2D images were azimuthally integrated and reduced to one-dimensional scatting profiles. DSC was used to determine the thermal stability of dehydrated samples. The measurement was conducted using TA Discovery DSC 2500 with 3 ~ 5 mg of samples and processed by Trios software, with the heating and cooling rate of 10 °C/min from room temperature to 250 °C. To monitor possible weight loss of the polymers and their composites, thermogravimetric analysis (TGA) measurements were conducted using a TA Instruments Discovery TGA equipped with an infrared furnace, auto-sampler, and gas delivery module. Nitrogen was used as the purge gas at a flow rate of 25 mL/min. Temperature ramp measurements were performed at 20 °C/min from ambient to 700 °C using standard platinum crucibles.

X-ray microimaging with phase contrast was used to see through the bulk samples and analyze the granular dispersion of starch. **Figures 2a, 2b, 2c** are X-ray micrographs of starch and water suspension, chitosan and water suspension, and starch-chitosan and water suspension, respectively. Randomly packed starch particles with a granular structure are clearly shown in **Figure 2a**. A Fast Fourier Transform reveals the most probable nearest particle distance is about 5.7 μm (data not shown), which is consistent with the nominal size of rice starch particles of 2 to 7 μm [22]. In stark contrast, the X-ray micrograph (**Figure 2b**) of chitosan in water shows a fiber-like structure of ~ 100-μm long and ~ 10-μm wide. Interestingly, when 9 wt% chitosan is mixed with 29 wt% starch in 62 wt% water, the relatively large chitosan microfiber bundles found in **Figure 2b** completely disappeared. Instead, the image is almost identical to that of the starch suspension (**Figure 2a**). The granular structure of starch is preserved, which signifies that the chitosan nanofibers have a morphologically significant difference from the original microbundle shape. They are now dispersed between the starch granules. Essentially, the absence of the chitosan microstructure indicates that chitosan microfibers have been 'dissolved' in a 'solvent' of a granular starch particle matrix. The size of the 'dissolved' chitosan structure is so small (less than the X-ray imaging resolution of 1 µm) and homogenous that chitosan became invisible in the X-ray micrographs. Here, with X-ray microimaging, we were able to observe a phenomenon where the chitosan particles underwent a morphological change due to the presence of starch granules, which allowed for the chitosan nanofibers to disperse between the granular suspension of starch particles.

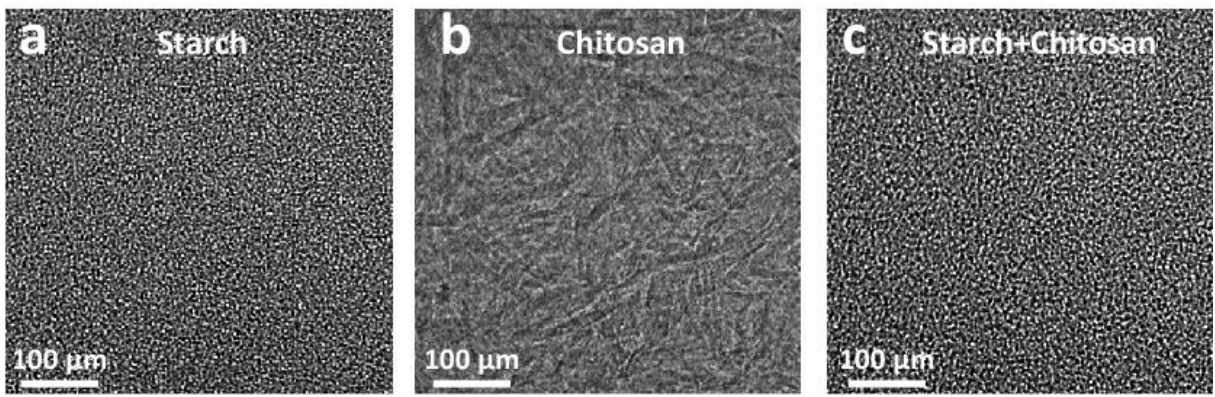

**Figure 2. X-ray micrographs reveals the dispersion of chitosan in a starch granule system. A.** X-ray microimage of a pure starch suspension, displaying an inter-particle distance of around 1 µm. **b.** X-ray microimage of a pure chitosan suspension, highlighting microbundles of length ~ 100 µm and width ~ 10 µm **c.** X-ray microimage of starch-chitosan suspension in water, highlighting how starch promotes the morphological change of chitosan from microbundles to nanofibers in suspensions.

To further investigate the sample's nanostructure after the 'solvation' of chitosan in starch, we performed SAXS measurements, as shown in **Figure 3**. The SAXS data of the suspension of chitosan alone exhibited a broad peak around the momentum transfer, $q_{max}$ = 0.55 nm$^{-1}$, indicating a weak order in the microfiber bundle seen in the X-ray micrograph (**Figure 2b**). The bundle spacing is about $\frac{2\pi}{q_{max}} \approx$ 11 nm. The SAXS data of starch shows a small but relatively sharp peak at 0.63 nm$^{-1}$, corresponding to a 10-nm spacing, typical to starch's microcrystalline phase in granular structure. As soon as chitosan is dissolved in starch, the chitosan broad peak completely disappears. The SAXS profile of the 'dissolved' sample overlaps the one for starch alone, indicating the disappearance of the chitosan microfiber bundles and, instead, the dispersion of chitosan nanofibers between starch granules.

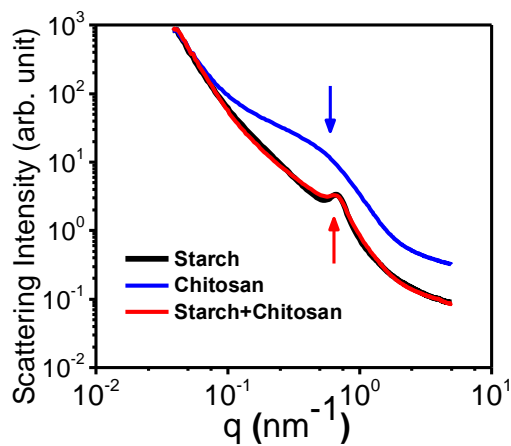

**Figure 3. SAXS profiles of starch, chitosan, and starch-chitosan samples.** The data displays the disappearance of the broad chitosan microbundle broad peak for the starch-chitosan sample at 0.55 nm$^{-1}$ and the prevalence of a microcrystalline peak at 0.63 nm$^{-1}$, proving a nanoscale morphological change.

The thermograph measured with DSC of dehydrated samples also supports the notion of 'dissolved' chitosan nanofibers. In **Figure 4a**, DSC scans of chitosan, starch, and starch-chitosan samples are presented in the temperature range of 150 to 250 °C, where all the thermal events occurred in the dehydrated samples. Chitosan alone shows a broad exothermic peak from 160 to 185 °C, with the peak position at 172 °C, indicating the structural change of the microfibers. The starch sample shows a relatively sharp exothermic peak at 182 °C, indicating the degradation of the granular structure. The starch-chitosan sample shows an elevated and more collaborative structural change that peaks at 187 °C. The peak is notably sharper than that of the 'pure' starch and 'pure' chitosan samples. To confirm that the peaks were caused by structural changes in the composite rather than mass loss of the materials, we also conducted TGA tests shown in **Figure 4b**. We observed that the weight of the material of all three samples stayed relatively constant in the temperature range 150 to 250 °C. None of the materials lost any mass; thus, the thermotransition peak of the starch-chitosan sample was caused by a structural change and possibly a bond rearrangement in the composite. The increased amount of energy introduced into the system indicated homogeneity of the composite and may have been caused by ionic interactions, a result of polyelectrolyte bonding between positively charged chitosan and negatively charged starch. Not only are the chitosan nanofibers evenly dispersed throughout starch, but they can also reinforce starch particle structure, promoting thermal stability.

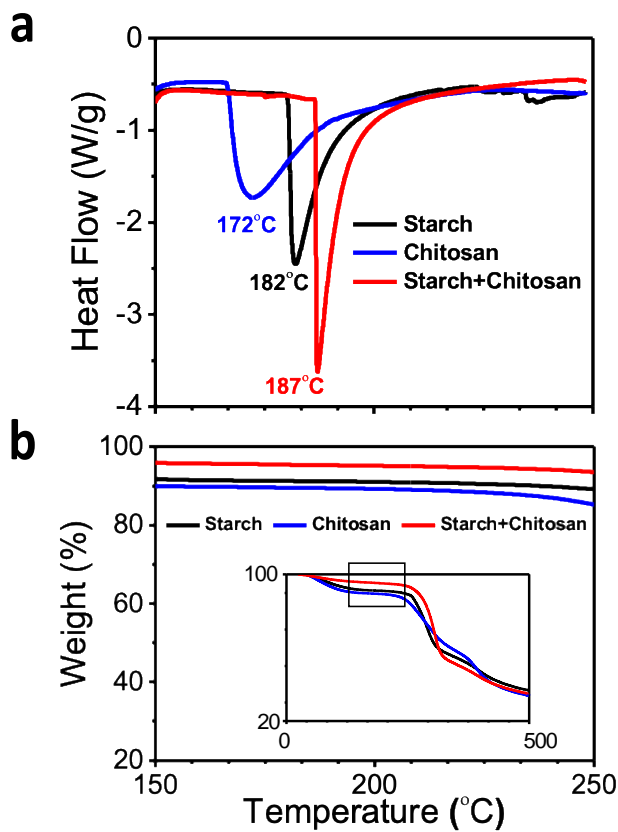

**Figure 4. Thermal stability analyses of starch-chitosan composites. a.** DSC thermograph in the range 150 to 250 °C of dehydrated samples of chitosan, starch, and starch-chitosan, with peaks at 172, 182, and 187 °C respectively. **b.** TGA tests of the chitosan, starch, and starch-chitosan samples in the temperature range of 150 to 250 °C with an inset of the whole TGA curves with a range from 0 to 500 °C. Relatively constant weight is evidenced through the range of 150 to 250 °C, proving that the DSC peaks are due to structural changes.

The dispersion of chitosan in the starch-chitosan composite was further investigated with SAXS of degraded samples. As shown in **Figure 5**, the SAXS profile of the degraded starch sample has no evident microcrystalline peak at 0.63 nm$^{-1}$, signaling the absence of starch particles. The chitosan sample in the same environment retained its broad peak at 0.55 nm$^{-1}$. Interestingly, the starch-chitosan composite sample does not have either peak, both the starch microcrystalline peak and the chitosan broad peak. These results prove that chitosan uniformly disperses in the granular starch matrix. If chitosan had not gone through a morphological change after dispersion, the chitosan broad peak should have reappeared once all the starch particles underwent degradation. However, even when the starch matrix disappeared after degradation, the non-existence of the broad chitosan microbundle pattern is evident. The absence of the chitosan broad peak, therefore, further confirms the homogeneity of the chitosan dispersion. It also signifies that the chitosan microbundles underwent a permanent structural change, with such an alteration enabled by starch granules.

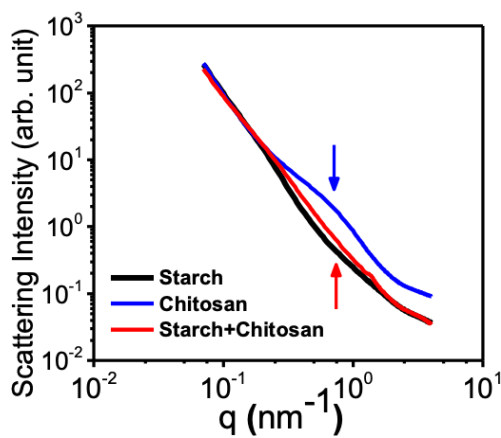

**Figure 5. SAXS profiles of starch, chitosan, and starch-chitosan samples after being in the same degrading environments.** The data shows the disappearance of both the broad chitosan microbundle broad peak at 0.55 nm$^{-1}$ and the microcrystalline peak at 0.63 nm$^{-1}$ for the starch-chitosan sample, proving permanent morphological change.

To summarize, by quantifying and qualifying the morphology of chitosan and starch on a mesoscale, we have shown the dispersion of chitosan in a starch matrix, which represents a novel solvation process of two biopolymers with similar monomer structures. Using X-ray microimaging, we visualized the granular starch suspension as well as the chitosan microbundles on a micrometer-length scale. By mixing the two materials, we could observe how the starch granules dissembled the chitosan microbundles into nanometer-scale structures, which were not detectable by X-ray imaging. This morphological change was further investigated through SAXS. The original broad peak from chitosan microbundles completely disappeared in the composite sample, and, instead, the only significant diffraction peak was the starch microcrystalline peak. DSC analysis advanced the idea of possible ionic interactions, as well as revealing a structural benefit of chitosan nanofibers dispersed between starch granules. Not only did we observe a morphological change in the composite, but also thermal, thus structural, stability that possibly contributed to the change. The mesoscale miscibility between starch and chitosan allows for an extremely efficient dispersion process of chitosan while maintaining its active functional properties at a low cost. The ability of chitosan to 'dissolve' in a suspension of starch, which enables the formation of a new composite material, has a variety of benefits. Not only is the composite biocompatible, but it also is possibly biodegradable, hemostatic, and antibacterial, among many other properties. This interaction between chitosan and starch may be especially applicable when creating further composite materials. These water-based materials are usually unsuitable for chitosan alone, as it cannot be dissolved in such a material, but the solvation of starch allows it to be viable. An external polymer structure can provide a mechanical matrix for starch granules and chitosan nanofibers, allowing the two to exhibit its biocompatible, active properties while maintaining structural integrity. Thus, with the potential mechanical strength, hemostatic properties, and biocompatibility, this material could be suitable, for example, for active drug delivery or as a hemostat for surgeries. This unique, low-cost, and simple mesoscale interaction will have a significant impact on chitosan's biological, chemical, and physical applications.

**Acknowledgment**

This work was supported by the National Science Foundation (NSF MRSEC, DMR 1420709). This research used resources at the APS, a U.S. Department of Energy (DOE) Office of Science User Facility operated for the DOE Office of Science by Argonne National Laboratory under Contract No. DE-AC02-06CH11357. We also thank Xiaobing Zuo, Qing Zhang, Ya Gao, Miaqi Chu, Jin Wang, and other beamline staff at Sectors 7, 8, and 12 of the APS, for providing technical support to the collection of SAXS, x-ray microimaging data.

## Competing financial interests

The authors declare no competing financial interests.